\documentclass[twocolumn]{aastex631}
\usepackage{makecell}
\usepackage{amsmath}

\begin{document}

\title{Supermassive primordial black holes for the GHZ9
and UHZ1 observed by the JWST}

\author[0009-0003-1218-2569]{Hai-Long Huang}
\affiliation{School of Fundamental Physics and Mathematical
    Sciences, Hangzhou Institute for Advanced Study, UCAS, Hangzhou
    310024, China}
\affiliation{School of Physical Sciences, University of
Chinese Academy of Sciences, Beijing 100049, China}
\email{huanghailong18@mails.ucas.ac.cn}

\author{Yu-Tong Wang}
\affiliation{School of Fundamental Physics and Mathematical
    Sciences, Hangzhou Institute for Advanced Study, UCAS, Hangzhou
    310024, China}
\affiliation{School of Physical Sciences, University of
Chinese Academy of Sciences, Beijing 100049, China}
\email{wangyutong@ucas.ac.cn}

\author{Yun-Song Piao}
\affiliation{School of Fundamental Physics and Mathematical
    Sciences, Hangzhou Institute for Advanced Study, UCAS, Hangzhou
    310024, China}
\affiliation{School of Physical Sciences, University of
Chinese Academy of Sciences, Beijing 100049, China}
\affiliation{International Center for Theoretical Physics
    Asia-Pacific, Beijing/Hangzhou, China}
\affiliation{Institute of Theoretical Physics, Chinese
    Academy of Sciences, P.O. Box 2735, Beijing 100190, China}
\email{yspiao@ucas.ac.cn}

\begin{abstract}
The high redshift ($z>10$) galaxies GHZ9 and UHZ1 observed by the
James Webb Space Telescope (JWST) are very massive and have
exceptionally high black hole-to-star mass ratios with the central
black hole masses $M\gtrsim 10^7\rm~M_\odot$. In this paper, we
explore the possibility that they are seeded by the supermassive
primordial black holes (SMPBHs), which came into being in the very
early universe, with initial masses $\sim 10^7\rm~M_\odot$. We
present the self-similar accretion solutions for SMPBHs, and find
that the mass growth of SMPBHs during pregalactic era may be
negligible. These SMPBHs, when the redshift $z\lesssim 20$, can
accelerate seeding high-redshift galaxies and their baryonic
content, and consequently explain the central supermassive black
holes (SMBHs) of high-redshift massive galaxies through
sub-Eddington accretion. According to our results, SMPBHs actually
could lead to the existence of more massive SMBHs at higher
redshifts compared to other SMBH seed scenarios, specially SMBHs
with masses $M\gtrsim 10^7~\rm M_\odot$ at $z>20$ might
only origin from SMPBHs, thus the corresponding observation can
serve as a potential probe to PBHs.
\end{abstract}


\tableofcontents

\section{Introduction} \label{sec:intro}

Recent observations with the JWST have identified a population of
quasars powered by SMBHs already in place in the first few hundred
million years after Big Bang \citep[e.g.][]{2023ApJ...953L..29L,
2023arXiv230605448M,2023A&A...677A.145U,
2023ApJ...955L..24G,2023ApJ...957L...7K,2023arXiv231203589U,
2023arXiv230805735F,2023arXiv230512492M,2024ApJ...960L...1N,
2024ApJ...965L..21K,2024NatAs...8..126B}. These comprise the
so-called ``Little red dots", which are compact and heavily
obscured, characterised by a steep red continuum in the rest-frame
UV
\citep{2023ApJ...959...39H,2023ApJ...954L...4K,2023arXiv230801230M,
2024ApJ...964...39G,2023arXiv231203065K,2024ApJ...968...38K,
2024ApJ...963..129M,2024arXiv240810305G,2024arXiv240811890A}.
Generating these SMBHs at very high redshifts is challenging due
to the lack of cosmic time to assemble within the standard
cosmological model. Therefore, it is crucial to understand the
nature of the SMBHs' ancestor and the processes governing their
mass growth. Many possible astrophysical scenarios for the
formation of SMBH seeds have been proposed in the literature so
far, including (1) light seeds ($\sim10^{1-2}\rm~M_\odot$,
$z\sim20-30$) from the first generation of metal-free (Population
\uppercase\expandafter{\romannumeral3}) stars
\citep{2001ApJ...551L..27M,2013RPPh...76k2901B,
2014ApJ...781...60H,2015ApJ...814...18H,2016ApJ...824..119H,
2020ApJ...892L..14S}; (2) intermediate seeds ($\sim10^{3-4}
\rm~M_\odot$, $z\sim10-20$) from the collapse of supermassive
stars or runaway collisions in dense, nuclear star clusters
\citep{2012MNRAS.421.1465D,2020ARA&A..58..257G}; (3) heavy seeds
($\sim10^{5-6}\rm~M_\odot$, $z\sim10-20$) produced by the direct
collapse of pristine, massive gas clouds or pre-galactic disks
\citep{1994ApJ...432...52L,
2006MNRAS.371.1813L,2006MNRAS.370..289B,2007MNRAS.377L..64L,
2023arXiv231213837I,2024arXiv240218773J}.

\begin{table*}
\centering
\begin{tabular}{lccccc}
\hline
 & Redshift & SMBH mass & Star formation rate & Stellar mass & Reference\\
\hline GHZ9 & $10.37_{-1.09}^{+0.32}$ &
$8.0_{-3.2}^{+3.7}\times10^7\rm~M_\odot$ &
$0.56_{-0.29}^{+0.23}~\rm~M_\odot~{\rm yr}^{-1}$ &
$4.9_{-2.6}^{+2.0}\times10^7\rm~M_\odot$ &
\citet{2024ApJ...965L..21K} \\
UHZ1 & $10.3_{-1.3}^{+0.6}$ &
$4.0_{-3.0}^{+6.0}\times10^7\rm~M_\odot$ & $4.4~\rm~M_\odot~{\rm
yr}^{-1}$ & $0.4_{-0.2}^{+1.9}\times10^8\rm~M_\odot$ &
\citet{2024NatAs...8..126B} \\
\hline
\end{tabular}
\caption{Derived physical properties for GHZ9 and UHZ1. } \label{tab:tab1}
\end{table*}

However, such explanations do not always suffice.
Among the SMBHs observed by JWST, GHZ9 \citep{2024ApJ...965L..21K} and
UHZ1 \citep{2024NatAs...8..126B} stand out as particularly massive
and at high redshift, as summarized in Table~\ref{tab:tab1}.
If one insists on light black hole (BH) seeds, unrealistically
sustained super-Eddington accretion rates are required. Heavy BH
seeds are the most promising scenarios to explain the existence of
these SMBHs without invoking super-Eddington accretion
rates. However, their formation rate is yet to be established
\citep[e.g.][]{2017PASA...34...31V,2020ARA&A..58...27I} and their
growth efficiency might depend on the formation site within the
host galaxy \citep[e.g.][]{2021MNRAS.502..700C,2024arXiv240614658B}.
There are
other possible explanations, such as supermassive dark stars
\citep{2023arXiv231213837I}, exotic objects
\citep{2024arXiv240313068I}, enhanced gravitational forces in
modified theories of gravity \citep{2024EPJC...84..238Z}, and
low-mass PBH seeds accompanied by Eddington-limited accretion
\citep{2002PhRvD..66f3505B,2004PhRvD..70f4015D,2015PhRvD..92b3524C,
2006CQGra..23.1875K,2021MNRAS.501.2029C,2023arXiv230309391Y,
2023arXiv231000671D,2024PhR..1054....1C,2024arXiv240707162D},
among others \citep[e.g.][]{2023arXiv231214738C,2024arXiv240715781H}.

It is intriguing to consider the possibility that SMBHs can
naturally come into being in very early universe prior to galaxies
themselves, potentially making them primordial (referred to as
primordial black holes, PBHs
\citep{1974MNRAS.168..399C,Zeldovich:1967lct,1971MNRAS.152...75H,
1975ApJ...201....1C,1975Natur.253..251C,1993PhRvD..48..543C,
1994PhRvD..50.7173I,1996PhRvD..54.6040G,1996NuPhB.472..377R,
2019PhRvL.123g1102M,2016PhRvL.116t1301B,2016PhRvL.117f1101S}),
since their abundance is closely related to the physics of very
early universe, and independent of the host galaxy. In this paper,
we investigate whether GHZ9 and UHZ1 can be seeded by the SMPBHs
with initial masses $\sim 10^7\rm~M_\odot$. In
Section~\ref{sec:2}, we describe the origin of SMPBHs.
Section~\ref{sec:3} presents the self-similar accretion model for
SMPBHs and its implication for the growth of SMPBHs during
pregalaxic era. In Section \ref{sec:seed}, we show how such SMPBHs
can seed high-redshift galaxies like GHZ9 and UHZ1. We draw
concluding remarks in Section~\ref{sec:results}.






\section{SMPBHs before matter-radiation equality} \label{sec:2}

In the standard picture, PBHs sourced by the collapse of large
density fluctuations in the radiation-dominated era
\citep{2016PhRvD..94h3504C,2017JPhCS.840a2032G,2020ARNPS..70..355C,
2021RPPh...84k6902C,2021JPhG...48d3001G,2021arXiv211002821C,
2022arXiv221105767E,2018CQGra..35f3001S,2024CQGra..41n3001D}.
The typical mass of a PBH is approximately equal to or
smaller than the mass within the particle horizon at the redshift
of its formation, $z_{\rm for}$ \citep[e.g.][]{2024arXiv240605736C}:
\begin{equation}
    M_{\rm for}\simeq \gamma\left(\frac{1+z_{\rm eq}}{1+z_
    {\rm for}}\right)^2M_{\rm eq}\mbox{\ .}
\end{equation}
Here, $M_{\rm eq}\simeq3.1\times10^{16}\rm~M_\odot$ is the mass of
the horizon at the redshift of matter-radiation equality
$z_{\text{eq}}\approx3400$, and $\gamma\le1$ is the ratio of the
mass of the PBH to the mass of the particle horizon at $z_{\rm
for}$. Hence, the mass range of a PBH can extend from
$10^{-18}\rm~M_\odot$ to $10^{16}\rm~M_\odot$ in principle
\footnote{Those smaller than $\sim10^{-18}\rm~M_\odot$ would have
evaporated by now due to Hawking radiation, or see recent
\citet{2024arXiv240902804C,2024arXiv240902807C}.}, although
current observations indicate that SMBHs in galactic nuclei
typically have masses extending up to only nearly
$10^{11}\rm~M_\odot$ \citep{2011Natur.480..215M}. It is worth
mentioning that producing SMPBH with masses above
$10^6\rm~M_\odot$ is strictly constrained by CMB $\mu$-distortion
observations \citep{2014PhRvD..90h3514K,2023PhRvL.130q1401D,
2018PhRvD..97d3525N,2021PhRvD.104f3526D}. To overcome these
constraints, significant non-Gaussianity must be invoked
\citep{2021PhRvD.103f3519A,
2023EL....14249001G,2024JCAP...04..021H,2024arXiv240418475B,
2024arXiv240418474S,2024PhRvL.133b1001C,2024arXiv240620089I} or
alternative mechanisms, such as PBHs origin from the
supercritical bubbles nucleated during inflation (which are not
constrained by $\mu$-distortion, e.g. \citet{2024arXiv240418475B}),
must be considered. In the latter case, the resulting SMPBHs might
have a multiple-peaks mass spectrum
\citep{2023arXiv230617577H,2024PhRvD.110b3501H} and the mass
function might extend up to $\sim10^{16}\rm~M_\odot$
\citep{2016JCAP...02..064G}.

During the radiation-dominated era ($z_{\rm for}>z>z_{\text{eq}}$),
it has been shown that there is very little accretion
\citep{1974MNRAS.168..399C}, hence we can approximately assume that
$M^{\rm ini}\equiv M(z_{\rm eq})=M_{\rm for}$.

\section{Accretion of SMPBHs} \label{sec:3}




\subsection{Growth of PBHs by accretion}

The accretion of PBHs in the intergalactic medium may
significantly alter their masses in principle. This phenomenon was
initially investigated by \citet{1981MNRAS.194..639C}, who assumed
that PBHs accrete at the Bondi rate. Subsequent studies
\citep[e.g.][]{2008ApJ...680..829R,2007ApJ...662...53R,
2017PhRvD..95d3534A,2023arXiv231214097D} have further explored PBH
accretion.

Ignoring the effect of mergers (see Appendix~\ref{subsec:mergers}), we
have
\begin{equation} \label{eq:31}
    \dot{M}=(1-\epsilon)\dot{M}_{\text{accr}}\mbox{\ .}
\end{equation}
Here, we assume that accreted mass is converted into outgoing
radiation with the radiative efficiency $\epsilon$. Typically,
$\epsilon=0.125$ is adopted \citep[e.g.][]{2017MNRAS.467.3475N}.
The associated bolometric luminosity is
$L=\epsilon\dot{M}_{\text{accr}}c^2$, where $c$ is the speed of light.
The dimensionless accretion rate is defined as
\begin{equation} \label{eq:32}
    \dot{m}\equiv\frac{\dot{M}_{\text{accr}}}{\dot{M}_
    {\text{Edd}}}\mbox{\ ,}
\end{equation}
in terms of the Eddington accretion rate $\dot{M}_{\text{Edd}}
\equiv L_{\text{Edd}}/(\epsilon c^2)$ with the Eddingtom
luminosity \citep{eddington1926book}
\begin{equation} \label{eq:6}
    L_{\text{Edd}}=\frac{4\pi GMm_pc}{\sigma_T}\approx1.3
    \times10^{38}\left(\frac{M}{\rm~M_\odot}\right)
    ~\textrm{erg}~\textrm{s}^{-1}\mbox{\ ,}
\end{equation}
where $G$ is the universal gravitational constant, $m_p$ is
the proton mass and $\sigma_T$ is the Thomson scattering cross-section.
Thus for a PBH with an initial mass $M^{\rm ini}=M(z_{\rm eq})$,
we can obtain the evolution of its mass by solving the equation
\begin{equation} \label{eq:34}
    \frac{\text{d}M(M,t)}{\text{d}t}=(1-\epsilon)
    \dot{M}_{\text{Edd}}\dot{m}\mbox{\ ,}\quad
    z(t)\in\left[z_{\text{eq}},z_{\text{cut}}\right]\mbox{\ ,}
\end{equation}
where $z_{\text{cut}}\sim\mathcal{O}(10)$ is the effective cut-off redshift,
noting the significant uncertainties in modeling the accretion rate at
relatively small redshifts \citep{2023arXiv231214097D}. It is
argued in \citet{2020JCAP...06..044D,2021PhRvD.103b3026W} that PBH
accretion can be considered negligible below $z_{\text{cut}}$.
This is due to the increased velocities of PBHs as they fall into
the gravitational potential wells of large-scale structures, which
suppress their accretion \citep{2020JCAP...07..022H}.


\subsection{Self-similar accretion of SMPBHs during pregalactic era}

The simplified basic equation under the steady-state approximation
can be found in \cite[e.g.][]{2023arXiv231214097D}. The
cosmological spherical accretion solutions described by the Bondi
solutions are \citep{bondi1952,hoyle1939,bondi1944}:
\begin{equation} \label{eq:bondi}
    \dot{M}_{\text{bondi}}(M,z)=4\pi\lambda\rho_{\text{gas}}v_
    {\text{eff}}r_B^2\mbox{\ ,}
\end{equation}
where $r_B=GM/v_{\text{eff}}^2$ is the Bondi-Hoyle radius in terms
of the effect velocity $v_{\text{eff}}=\sqrt{v_{\text
{rel}}^2+c_s^2}$, $v_{\text{rel}}$ is the velocity with respect to
the surrounding hydrogen gas with the density $\rho_{\text{gas}}$
and $c_s$ is the sound speed. The analytical expressions of
$\rho_{\text{gas}}$ and $c_s$ can be found in
\citet{2007ApJ...662...53R} and are reported in
Appendix~\ref{sec:appendixA}. For the accretion parameter
$\lambda$ one can refer to \citet{2017PhRvD..95d3534A}, which takes
into account its suppression by the Compton drag and Compton
cooling by CMB photons.



However, the Bondi solutions are valid only when the sound crossing
time is less than the Hubble time $t_{\rm cr}\sim r_B/c_{s,\infty}
<t_H$, or equivalently,
\begin{equation}
    M\lesssim10^4~\rm M_\odot,
\end{equation}
where
$c_{s,\infty}$ is the sound speed at infinity. Thus the Bondi
solutions are not applicable for SMPBHs
with masses $M\gtrsim 10^6~\rm M_\odot$.



Therefore, it is necessary to develop the accretion formulas
suitable to SMPBHs. It is reasonable to consider a spherically
symmetric flow in the gravitational field of a central mass $M$
under the Newtonian approximation \citep{2008ApJ...680..829R,
2020JCAP...04..052D,2020JCAP...06..044D}. In the spherical
coordinates $(r,\theta,\phi)$, the basic equations for accretion
are the continuity equation (mass equation)
\begin{equation} \label{eq:continuity}
    \frac{\partial\rho}{\partial t}+\frac{1}{r^2}\frac
    {\partial}{\partial r}(r^2\rho v)=0,
\end{equation}
and the equation of motion in the radial direction (momentum
equation) \begin{equation} \label{eq:momentum}
    \frac{\partial v}{\partial t}+v\frac{\partial v}{\partial r}
    =-\frac{1}{\rho}\frac{\partial p}{\partial r}-\frac{GM}{r^2}-\beta v,
\end{equation}
where $\rho$ is the density of surrounding gas, $p$ is its
pressure, $v$ is the radial infalling velocity of gas particles,
and
\begin{equation} \label{eq:beta}
\beta v =\frac{4}{3}\frac{\chi_e\sigma_T\rho_{\rm CMB}}{m_pc}v+Hv
\end{equation}
is the external drag force including the Compton drag ($\chi_e$ is
the electron fraction, $\rho_{\rm CMB}$ is the CMB energy density
and $H$ is the Hubble parameter) and the Hubble term assumed
spatially constant but time-dependent. As discussed in
\citet{1995PASJ...47...73T}, the pressure term in
Equ.~\eqref{eq:momentum} is smaller than the other terms assuming
a power-low density profile $\rho=r^{-\alpha}$ with
$\alpha\sim\mathcal{O}(1)$, and thus will be been neglected
in the following analysis.

We solve Equs.~\eqref{eq:continuity} and \eqref{eq:momentum} at any
redshift (or cosmic time $t$) by self-similar transformations
techniques developed in
\citet{1995PASJ...47...73T,1974Ap&SS..26..183S,1984PASJ...36...87F}.
The method is outlined as follows.

The similarity variable $\xi$ is introduced as
\begin{equation}
    \xi=(GM)^{-1/3}rt^{-2/3}.
\end{equation}
The variables $(v, \rho)$ are transformed into the similarity
variables $(V, D)$ via
\begin{align}
    v&=\frac{r}{t}V(\xi)=(GM)^{1/3}t^{-1/3}\xi V(\xi),
    \\
    \rho&=r^{-\nu}D(\xi)=(GM)^{-\nu/3}t^{-2\nu/3}\xi^{-\nu}D(\xi),
    \label{eq:rhotoD}
\end{align}
where $\nu$ is a constant parameter which specifies the spatial
density distribution far away from the center.
The quantities $\xi$, $\xi V$ and $D$ correspond to the radius,
the velocity and the non-dimensional density at $t=1$ in
units of $GM=1$.

It is important to note that we take into account the
non-negligible viscous term
resulting from the Hubble expansion. This is reflected in the
calculation of the coefficient $\beta$, which should vary as
\begin{equation}
    \beta=\frac{\eta}{t}
\end{equation}
due to the requirement of dimension, with $\eta$ a
positive constant. In order to match its value with that given
by Equ.~\eqref{eq:beta}, we have
\begin{equation}
    \eta\simeq8.96\times10^{-6}\chi_e(1+z)^{5/2} + \frac{2}{3},
\end{equation}
where we use the fact that in the matter dominated epoch,
$t\propto(1+z)^{-3/2}$ and $H\sim2/(3t)$.
The ionization degree should change as $(1+z)^{-5/2}$ and we
choose $\eta=0.7$ to make $10^{-5}<\chi_e<1$ during the epoch
between $z_{\rm eq}$ and $z_{\rm cut}$.

After some manipulations, Equs.~\eqref{eq:continuity}
and \eqref{eq:momentum} are transformed as the following set
of differential equations:
\begin{align}
    \frac{{\rm d}D}{{\rm d}\xi}&=-\frac{(3-\nu)V+\xi({\rm d}V/
    {\rm d}\xi)}{\xi(V-2/3)}D, \label{eq:22}
    \\
    \frac{{\rm d}V}{{\rm d}\xi}&=-\frac{V^2+(\eta-1)V+\xi^{-3}}
    {\xi(V-2/3)}. \label{eq:23}
\end{align}
The boundary conditions at infinity ($\xi\to\infty$, or
$r\to\infty$) are
\begin{equation} \label{eq:boundary1}
    \xi V\to-\frac{1}{\eta+1}\xi^{-2}\quad {\rm or} \quad
    v\to-\frac{GM}{\eta+1}\frac{t}{r^2},
\end{equation}
and
\begin{equation} \label{eq:boundary2}
    D(\xi)\to C_1 \quad {\rm or} \quad \rho\to C_1r^{-\nu}.
\end{equation}
They are imposed in order to satisfy the realistic asymptotic
behavior. To ensure that $\rho$ remains constant spatially at
infinity, we must set $\nu=0$ according to Equ.~\eqref{eq:boundary2}.
The spatial constant $C_1$ is normalized as the mean cosmic gas
density $\rho_{\rm gas}\propto (1+z)^3$ given by Equ.~\eqref{eq:rhogas},
such that $\rho\to C_1=\rho_{\rm gas}$ as $r\to\infty$.

On the other hand, the boundary conditions in the center
region ($\xi\to0$, or $r\to0$) are given by
\begin{equation} \label{eq:boundary01}
    \xi V\to-\sqrt{2}\xi^{-1/2}\quad {\rm or} \quad v\to
    -\sqrt{2GM/r},
\end{equation}
and
\begin{equation} \label{eq:boundary02}
    D\to C_2\xi^{-3/2}\quad {\rm or} \quad \rho\to C_2\sqrt
    {GM}tr^{-3/2},
\end{equation}
where $C_2$ is a normalization constant similar to $C_1$.

The mass accretion rate is given by $\dot{M}=-4\pi r^2v\rho$.
According to the boundary conditions
Equs.~\eqref{eq:boundary01} and \eqref{eq:boundary02},
the mass accretion ratio at the center $r\to0$ is given by
\begin{equation} \label{eq:self_similar}
    \dot{M}_{\rm self-simlilar}(M,z)=4\sqrt{2}\pi GMC_2t.
\end{equation}
This will be used as the accretion rate of the central black hole
in the following analysis, see also \cite[e.g.][]{1995PASJ...47...73T}.


It is significantly noted that since the mass accretion rate for
the self-similar solutions Equ.~\eqref{eq:self_similar} is
proportional to $M$, considering Eddington accretion rate
Equ.~\eqref{eq:32}, we have the dimensionless accretion rate 
\begin{equation} \label{eq:mass-independent}
    \dot{m}=\frac{\dot{M}_{\rm
self-simlilar}}{\dot{M}_ {\text{Edd}}}=\left(\frac{\sqrt{2}\epsilon
c\sigma_T}{m_p}\right)C_2t\propto (1+z)^{3/2},
\end{equation}
independent of the mass of PBHs.







\begin{figure}[ht!]
    \includegraphics[width=\columnwidth,clip]{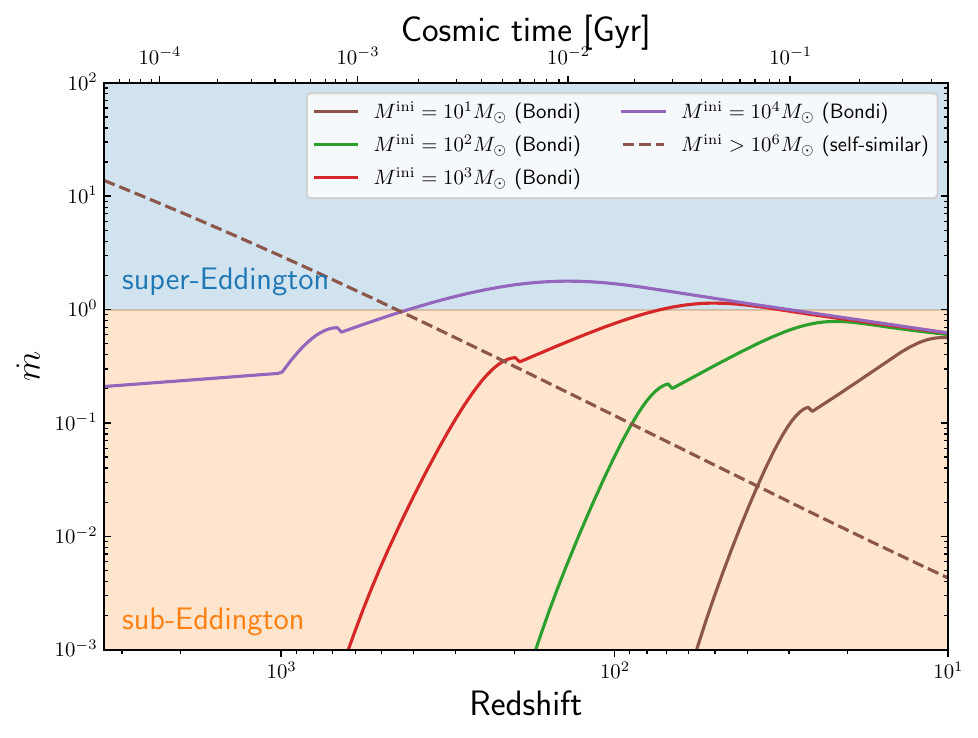}
    \caption{The dimensionless accretion rate $\dot{m}$ of pregalactic
    gas onto PBHs with different initial masses $M^{\rm ini}$ with respect to
    the redshift or cosmic time, calculated by Bondi solutions
    (Equ.~\eqref{eq:bondi}, solid lines) and self-similar
    solutions (Equ.~\eqref{eq:self_similar} for SMPBHs, dashed lines).
    Here, $\dot{m}$ is mass-dependent for the Bondi solutions, while it is
    independent of the mass for the self-similar
    solutions, and decay with the lowering of the
redshift, see Equ.~\eqref{eq:mass-independent}.}
    \label{fig:dotm}
\end{figure}

\begin{figure}[ht!]
    \includegraphics[width=\columnwidth,clip]{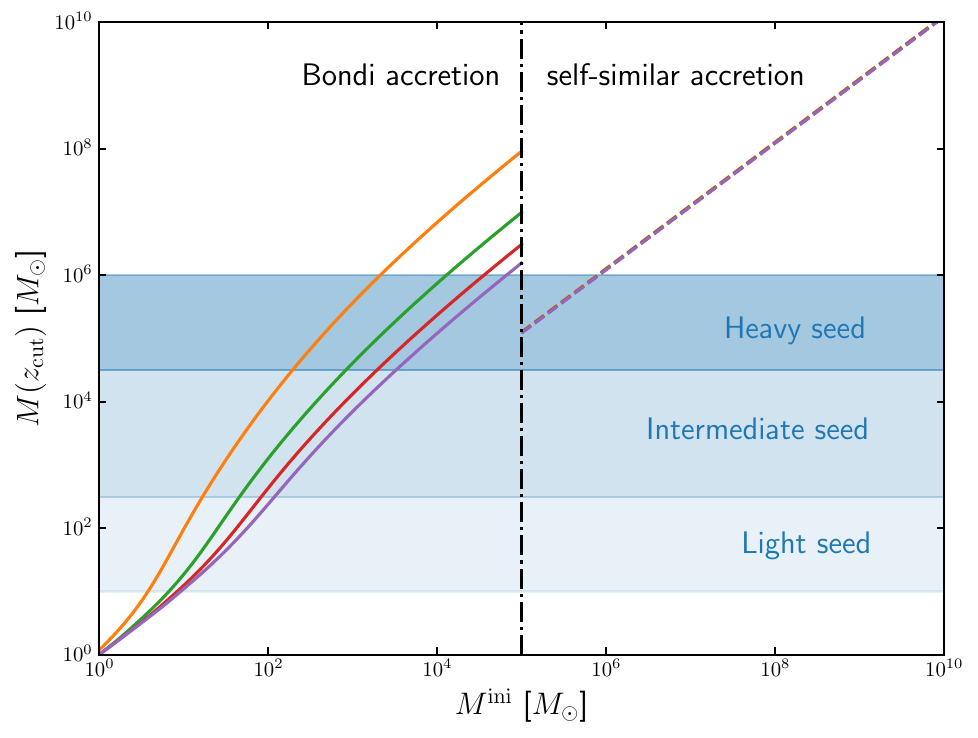}
    \caption{The masses $M(z_{\rm cut})$ of PBHs at
    $z_{\rm cut}$
    for different initial $M^{\rm ini}$ according to
    Equ.~\eqref{eq:34}, assuming different
    values for the cut-off redshift $z_{\rm cut}$= 10 (orange),
    15 (green), 20 (red) and 25 (purple).
As a comparison, the mass ranges of astrophysical SMBH seeds are
indicated by
    shaded regions.
    It shows that in the case of self-similar accretion
    the value of $z_{\rm cut}$ plays a negligible role on
    $M(z_{\rm cut})$.
    }
    \label{fig:mcut_mi}
\end{figure}

The self-similar accretion rate $\dot{m}$ for SMPBHs with respect
to the redshift or cosmic time is showed in Fig.~\ref{fig:dotm}.
The self-similar accretion results in SMPBHs undergoing a period
of super-Eddington accretion, which then shifts to sub-Eddington
accretion at redshifts $z\sim\mathcal{O}(10^2-10^3)$. Thus compare
to the Bondi accretion of PBHs with the masses $M^{\rm
ini}\lesssim10^4\rm~M_\odot$, the accretion of SMPBHs during
pregalactic era has negligible effect on the mass of SMPBHs, see
also Fig.~\ref{fig:mcut_mi}. However, this does not imply that the
actual accretion rate $\dot{M}_{\rm accr}$ of SMPBH is inherently
smaller (see Fig.~\ref{fig:accr_vs_merger} in
Appendix~\ref{subsec:mergers}).




\section{Seeding early galaxies like GHZ9 and UHZ1} \label{sec:seed}

In the scenarios where SMPBHs seed galaxies, they are located in
galactic nuclei, where they can accrete local gas and stars and
potentially merge with other galaxies
\citep{2010A&ARv..18..279V,2016PASA...33....7J,2020ARA&A..58...27I}.
Conversely, if SMPBHs are situated outside galaxies or galaxy
clusters, they would primarily accrete intergalactic gas
\citep{2021MNRAS.501.2029C}. Their mass growth is also governed by
Equ.~\eqref{eq:31}, but requires reassessment of $\dot{M}_{\rm
accr}$. Besides, it is regulated by the
processes of star formation and mechanical feedback, which both
lead to a depletion of gas inside the host galaxy
\citep{2022MNRAS.511..616T}.

\begin{figure*}[htb!]
    \centering
    \includegraphics[width=\textwidth]{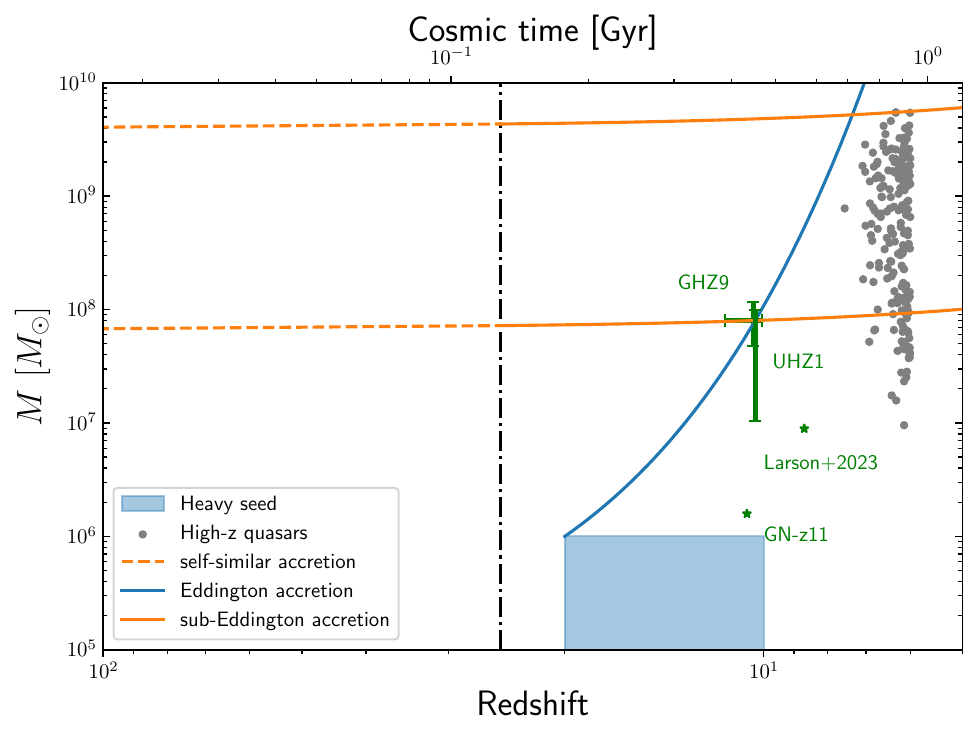}
\caption{The mass growth of SMPBH with respect to the redshift or
cosmic time. We use the dotted line to represent the growth of
PBHs during pregalactic era (here we choose
$z>z_{\text{cut}}=25$), and
    the solid blue (orange) line to depict the accretion of SMBHs
    in galactic nuclei at a rate of (0.02 times) the Eddington
    limit. Additionally, we display select JWST observations of
    high-redshift AGN \citep{2023ApJ...953L..29L,2024NatAs...8..126B,
    2023arXiv230801230M,2024ApJ...965L..21K} and 203 quasars at
    $z\ge6$ from \citet{2020ARA&A..58...27I}. Shaded regions indicate
    the initial BH seeds introduced in Section~\ref{sec:intro}.
    }
    \label{fig:total}
\end{figure*}

Here, we assume that the accretion of SMPBHs below $z_{\rm cut}$
is negligible ($\dot{m}\ll1$), \footnote{In fact, the
(super-)Eddington accretion onto seed PBHs to be evolved to SMBHs
should be constrained by observations of the high-redshift 21cm
line because of radiation emitted by the accretion disks and its
heating of the cosmological plasma. \citet{2022PhRvD.106d3539K}
showed that the mass of the corresponding seed PBHs should be
conservatively smaller than $10^6\rm~M_\odot$. We thank Kazunori
Kohri for bringing this point to us.}. Recent observations of
high-redshift active galactic nuclei suggest that most may be
dormant, exhibiting sub-Eddington accretion rates
\citep{2023MNRAS.524..176J}. In particular, 
\citet{2024arXiv240303872J} report a SMBH with mass of
$\sim4\times10^8\rm~M_\odot$ and accreting at a rate of only 0.02
times the Eddington limit (i.e. $\dot{m}=0.02$) at $z=6.68$.
According to Equs.~\eqref{eq:31}-\eqref{eq:34}, we have
\begin{equation} \label{eq:38}
    M(t)=M_{\rm cut}\exp\left[\frac{4\pi Gm_p\dot{m}}{c\sigma_T}\frac
    {1-\epsilon}{\epsilon}(t-t_{\rm cut})\right],
\end{equation}
where $M_{\rm cut}$ is the mass of SMPBH at $z_{\rm cut}$. In
Fig.~\ref{fig:total}, we find that both GHZ9 and UHZ1 can be
well explained as SMPBHs with such a sub-Eddington accretion rate.


It is known that GHZ9 and UHZ1 also show unprecedentedly high
black hole-to-stellar mass ratios compared to the local relations
\citep{1998AJ....115.2285M,2004ApJ...604L..89H,2013ARA&A..51..511K,
2015ApJ...813...82R}. In our case, a PBH of mass $M$ can bind a DM
halo with mass of \citep{2018MNRAS.478.3756C}
\begin{equation} \label{eq:35}
    M_h=\frac{1+z_{\rm eq}}{1+z}M
\end{equation}
at redshift $z$. The binding mass is derived from the fact that
PBH fluctuations grow as $(1+z)^{-1}$ during the matter-dominated
era, but this growth breaks down in the non-linear regime for
$z<z_{\rm cut}$. Then the mass growth rate of DM halos after
$z_{\rm cut}$ with the mass of $M_h(z_{\rm cut})=(1+z_{\rm
eq})/(1+z_{\rm cut})M$ is approximated as
\citep{2010MNRAS.406.2267F, 2022ApJ...938L..10I}
\begin{align} \label{eq:36}
    \dot{M}_h\simeq & ~46.1{\rm~M_\odot}~{\rm yr}^{-1}\left(\frac{M_h}
    {10^{12}\rm~M_\odot}\right)^{1.1} \notag \\
    &\times(1+1.11z)\sqrt{\Omega_m(1+z)^3+\Omega_\Lambda}.
\end{align}
If we allow the star formation to take place at $z_{\rm cut}$, we
have
\begin{align} \label{eq:37}
    M_*=
\begin{cases}
    0  \qquad  \qquad {\rm if}  \qquad z>z_{\rm cut},  \\
    \epsilon_* f_bM_h \quad  \ \, {\rm otherwise},
\end{cases}
\end{align}
where $f_b=\Omega_b/\Omega_m$ is the average fraction of baryons
in matter and $\epsilon_*$ is the star formation efficiency that
is assumed to be constant here.

\begin{figure*}
\centering
    {\includegraphics[width=8.5cm]{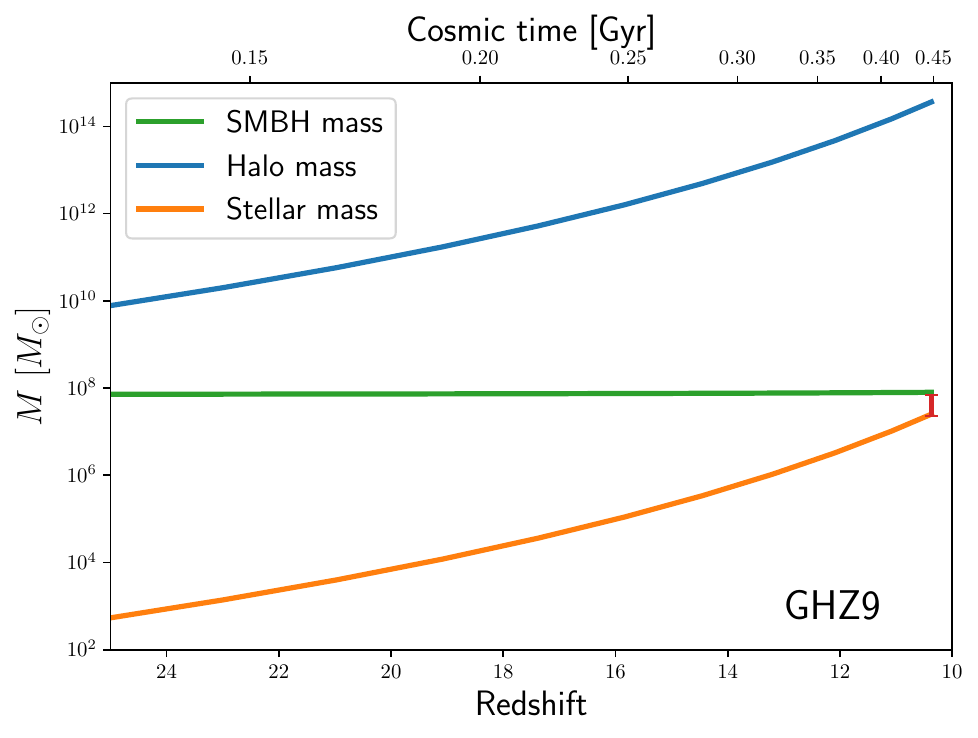}}
    \quad
    {\includegraphics[width=8.5cm]{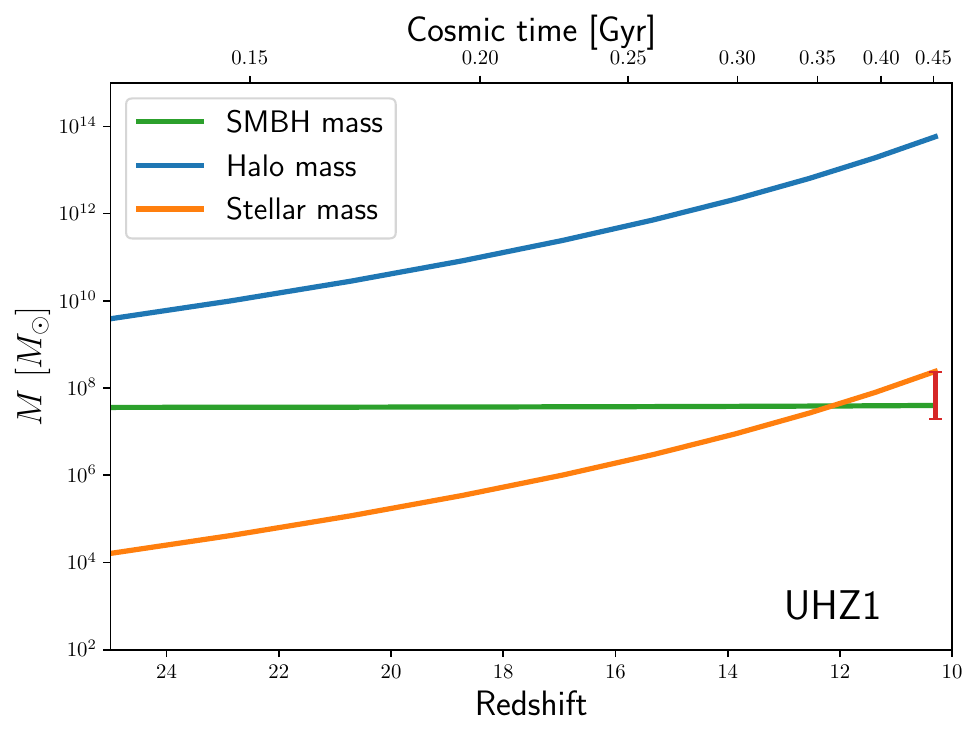}}
\caption{{\it Left panel:} Explaining GHZ9
($M=4\times10^7\rm~M_\odot$ and $z=10.37$) using SMPBH under the
assumption of sub-Eddington accretion ($\dot{m}=0.02$). Based on
the evolution of the SMBH mass derived with Equ.~\eqref{eq:38} and
the corresponding initial mass $M^{\rm ini}$, we compute the
evolution of the halo using Equs.~\eqref{eq:35} and \eqref{eq:36},
and subsequently the stellar mass using Equ.~\eqref{eq:37}. Here,
the star formation efficiency $\epsilon_*$ is determined by the
observed star formation rate listed in Table~\ref{tab:tab1}
through $\dot{M_*}=\epsilon_*f_b\dot{M_h}$. The stellar mass of
GHZ9 predicted by our SMPBH model (orange line) is in agreement
with the actual observed value, as indicated by the red error
bars. {\it Right panel:} Same as the left panel with the different
parameters for UHZ1.}
    \label{fig:stellar_mass}
\end{figure*}

The explanation of SMPBHs for GHZ9 and UHZ1 can naturally bring
required stellar masses, $2.52\times10^7\rm~M_\odot$ for GHZ9 and
$2.42\times10^8\rm~M_\odot$ for UHZ1, that consistent with
observations (red error bars), see Fig.~\ref{fig:stellar_mass}.




\section{Conclusions and discussion} \label{sec:results}


In this paper, we explore the possibility in which SMPBHs explain
the GHZ9 and UHZ1 observed by the JWST. We present the self-similar
accretion solutions for SMPBHs, and find that the mass growth of
SMPBHs during pregalactic era may be negligible. These SMPBHs
undergo modest self-similar accretion before galaxies form,
and could naturally account for high-redshift galaxies through
sub-Eddington accretion, see Fig.~\ref{fig:total}. According to
our results, SMPBHs actually could lead to the existence of more
massive SMBHs at higher redshifts compared to other SMBH seed
scenarios, specially SMBHs with masses $M\gtrsim 10^7~\rm
M_\odot$ at $z>20$ might only origin from SMPBHs, thus the
corresponding observation can serve as a potential probe to PBHs.
There might also exist an intergalactic population of SMPBHs that
is not directly related to the SMBHs observed in galactic nuclei.


Here, we define a cut-off redshift $z_{\rm cut}$ around the epoch
that the galaxies start to come into being, after which we neglect
the accretion artificially (only consider very low sub-Eddington
accretion). Though our results indicate that SMPBHs could explain
the over-abundance of SMBHs at high redshifts, the existence of
exceptionally massive BHs at very high redshifts and the
extraordinarily high black hole-to-stellar mass ratio revealed by
the JWST, numerical simulations are essential for studying the
co-evolution of SMBHs and their host galaxies in the non-linear
regime. It is known that understanding the co-evolution of nuclear
black holes and their host galaxies presents even greater
challenges, often necessitating complex and model-dependent
numerical simulations, see e.g.
\citet{2023arXiv231208422L,2024arXiv240113733B}. However,
it is interesting and significant to further investigate the
relevant issues.

\begin{acknowledgments}
We thank Dashuang Ye, Jun-Qian Jiang and Mingjun Liu for helpful
discussion. This work is supported by National Key Research and
Development Program of China, No. 2021YFC2203004, and NSFC,
No.12075246.
\end{acknowledgments}






\appendix

\section{The effect of mergers on the pregalactic growth} \label{subsec:mergers}

Initially, PBHs are randomly distributed in the early Universe
after their formation. If two PBHs happen to be separated by a
sufficiently short distance, they could decouple from the Hubble
flow and move towards each other due to gravitational attraction.
Subsequently, due to environmental disturbances, such as a third
neighboring PBH exerting a tidal torque, the two PBHs can avoid
a head-on collision and form a binary system. These binaries
will eventually merge due to gravitational radiation
\citep{1997ApJ...487L.139N}.

\subsection{Monochromatic mass function}

In this scenario, all PBHs have the same mass, and we assume
they undergo the same growth process. Thus, the mass function
is given by $\psi(m,z)\equiv\frac{m}{\rho_{\rm PBH}}\frac
{{\rm d}n}{{\rm d}m}=\delta(m-M(z))$. Here, $\rho_{\rm PBH}$
is the energy density of PBHs and $n(m)$ is the average number
density in the mass interval $(m,m+{\rm d}m)$.
Assuming that the merger rate per unit comoving volume $V$ per
unit time $t$ is $R(m,z)\equiv{\rm d}N_{\rm merger}/({\rm d}V
{\rm d}t)$, given by a specific PBH model (see Appendix~\ref{sec:appendixB}
for detail), the total merger
event number between redshift interval $(z,z+{\rm d}z)$ is
\begin{equation} \label{eq:3}
    {\rm d}N_{\rm merger}=R(M^{\rm ini},z)\frac{{\rm d}V_c}
    {{\rm d}z}\frac{{\rm d}t}{{\rm d}z}{\rm d} z,
\end{equation}
where ${\rm d}V_c/{\rm d}z=4\pi d_c^2/H(z)$ is the comoving volume
per unit redshift with the comoving distance $d_c$ \footnote{
Here, we do not consider the effect of the resolution of Hubble
tension on $H(z)$ since matter-radiation equality,
e.g. \citet{2020PhRvD.101h3507Y,2024arXiv240418579W},
which may be actually negligible.}. In Equ.~\eqref{eq:3}, we
assume for simplicity that the merger rate of PBHs is completely
determined by their initial properties (including initial mass
$M^{\rm ini}$, energy density $\rho^{\rm ini}_{\rm PBH}$ that are
not explicitly represented, their initial space distribution an so
on). Based on this assumption, the merger rate is simply a
function of redshift. Thus the average number of merging events
experienced by a single PBH between $(z,z+{\rm d} z)$ is
\begin{equation}
    2~{\rm d}N_{\rm merger}\left(n_{\rm PBH}\cdot\frac{{\rm d}
    V_c}{{\rm d}z}\right)^{-1}=\frac{2R(M^{\rm ini},z)}{n_{\rm
    PBH}}{\rm d}t,
\end{equation}
where $n_{\rm PBH}$ is the number density of PBHs and the factor
$2$ takes into account that every merge involves two PBHs.
We note that while the number density of PBHs is not affected
by the accretion process (as discussed in the next subsection),
it theoretically decreases as PBHs
merge. Here, we simply ignore this effect (which holds when the
merger rate is small) and assume that the number density is always
equal to  $\rho_{\rm PBH}^{\rm ini}/M^{\rm ini}$. We have
\begin{equation} \label{eq:dotmerger_mono}
    \dot{M}_{\rm merger}=\frac{{\rm d}M_{\rm merger}}{{\rm d}t}
    =\frac{2MR(M^{\rm ini},z)}{n_{\rm PBH}}.
\end{equation}
Then we turn to the calculation of $\dot{M}_{\rm merger}$ for
general mass functions.

\subsection{Extended mass function}

Ignoring the effect of the merger on the number density of
the PBHs as before, the evolution of the mass function is given by PBH
number density conservation, i.e. \footnote{Note the
difference between our definition of mass function $\psi(m)$
and that in \citet{2020JCAP...06..044D,2020JCAP...04..052D,
2020PhRvD.102d3505D}.}
\begin{equation}
    \frac{\rho_{\rm PBH}\psi(M(M^{\rm ini},z),z)}{M}{\rm d}M
    =\frac{\rho_{\rm PBH}^{\rm ini}\psi(M^{\rm ini},z^{\rm ini}
    )}{M^{\rm ini}}{\rm d}M^{\rm ini}.
\end{equation}
Mergers of PBHs do not change their energy density, i.e.
$\rho_{\rm PBH}=\rho_{\rm PBH}^{\rm ini}$.
We introduce a cross-grained discrete mass distribution
function, namely
\begin{equation}
    \int\psi(m,z){\rm d}m=1 \to\sum_i\psi_i\Delta=1,
\end{equation}
where $\psi(m,z)\to\psi_i$ is the binned mass distribution
function and ${\rm d}m\to\Delta$ denotes the resolution of
PBH mass. Denote the PBH mass of $i$-th bin is $m_i$, the
average number density of PBHs in the $i$-th bin is
\begin{equation}
    n_i\Delta=\frac{\rho_{\rm PBH}\psi_i\Delta}{m_i}.
\end{equation}

Similar to the previously used merger rate, the differential
merger rate adopted here $\mathcal{R}(m_i,m_j,z)\equiv{\rm d}N_{\rm merger}
/({\rm d}V{\rm d}t{\rm d}m_i{\rm d}m_j)$ is also given by the
initial condition of PBHs. We consider the mass growth rate of
the PBH in the $i$-th bin through mergers, i.e. $M^{\rm ini}
=m_i$. The total merger event number between redshift interval
$(z,z+{\rm d}z)$ for all PBHs in the $i$-th bin merging with PBHs
in the $j$-th bin is
\begin{equation}
    {\rm d}N_{\rm merger}=2\mathcal{R}(M^{\rm ini},m_j,z)\Delta^2
    \frac{{\rm d}V_c}{{\rm d}z}\frac{{\rm d}t}{{\rm d}z}{\rm d}z.
\end{equation}
The factor of 2 arises from from the fact that $\mathcal{R}(m_i,m_j,z)
=\mathcal{R}(m_j,m_i,z)$. Then
\begin{equation}
    \dot{M}_{\rm merger}=\frac{\sum_jm_j{\rm d}N_{\rm merger}}{{\rm d}t}\left(n_i\Delta\cdot\frac{{\rm d}V_c}{{\rm d}z}\right)^{-1}.
\end{equation}
As $\Delta\to0$, we have
\begin{equation} \label{eq:dotmerger_ex}
    \dot{M}_{\rm merger}=\frac{2M}{\rho_{\rm PBH}\psi(M,z)}\int m_j
    \mathcal{R}(M^{\rm ini},m_j,z){\rm d}m_j.
\end{equation}
One can confirm that this reduces to Equ.~\eqref{eq:dotmerger_mono}
if $\psi(m)=\delta(m-M(z))$, using $R(z)=\int\int\mathcal{R}(m_i,
m_j,z){\rm d}m_i{\rm d}m_j$.

For simplicity, we completely ignore the backreaction of PBH
growth on the merger rate (density). Specifically, we
independently examined mergers and accretion in the paper.
However, in reality, the merger process of a PBH binary will
affect the accretion of its components, and accretion, in turn,
influences the merger process. \footnote{For example, assuming
that PBHs undergo Bondi accretion (Equ.~\eqref{eq:bondi}),
accretion occurs on the binary as a whole when the orbital
separation is smaller than the Bondi radius of the binary (which
is defined by the total mass and the velocity of the center of
mass of the binary). In this configuration, the two binary
components accrete differently compared to isolated PBHs
\citep{2020JCAP...06..044D}. On the other hand, the mass growth of
PBHs will influence the semi-major axis of the binary in the
adiabatic approximation, altering the coalescence time and,
consequently, the merger rate (density)
\citep{2020JCAP...06..044D}.} In addition, we will ignore the
possibility that PBHs experience second-generation mergers, that
are coalescences in which a PBH binary might merger giving rise to
a new PBH, which later undergoes a further merger event with
another PBH
\citep{2019EPJC...79..717L,2020PhRvD.101h3008W,2020JCAP...04..052D}.

\subsection{Mergers vs. accretion}

Then we compare the sizes of $\dot{M}_{\rm accr}$ and
$\dot{M}_{\rm merger}$  for a PBH with masses $M=10\rm~M_\odot$
(described by the Bondi solutions) and $10^6\rm~M_\odot$
(self-similar solutions)
at different redshifts. The merger rate, detailed in
Appendix~\ref{sec:appendixB}, is highly dependent on the number density
of PBHs, which is tightly constrained by cosmological and
astrophysical observations \citep[e.g.][]{2018CQGra..35f3001S,
2021RPPh...84k6902C,2022arXiv221105767E,2023arXiv231019857B}.
For illustrative purposes, we consider a monochromatic PBH
population with energy density fractions of
$f_{\rm PBH}\equiv\rho_{\rm PBH}/\rho_{\rm DM}=10^{-3}$ and
$10^{-4}$ for both cases. Note that for a given energy density
$f_{\rm PBH}$, the number density of PBHs is inversely
proportional to their masses.

\begin{figure*}
\centering
    {\includegraphics[width=8.5cm]{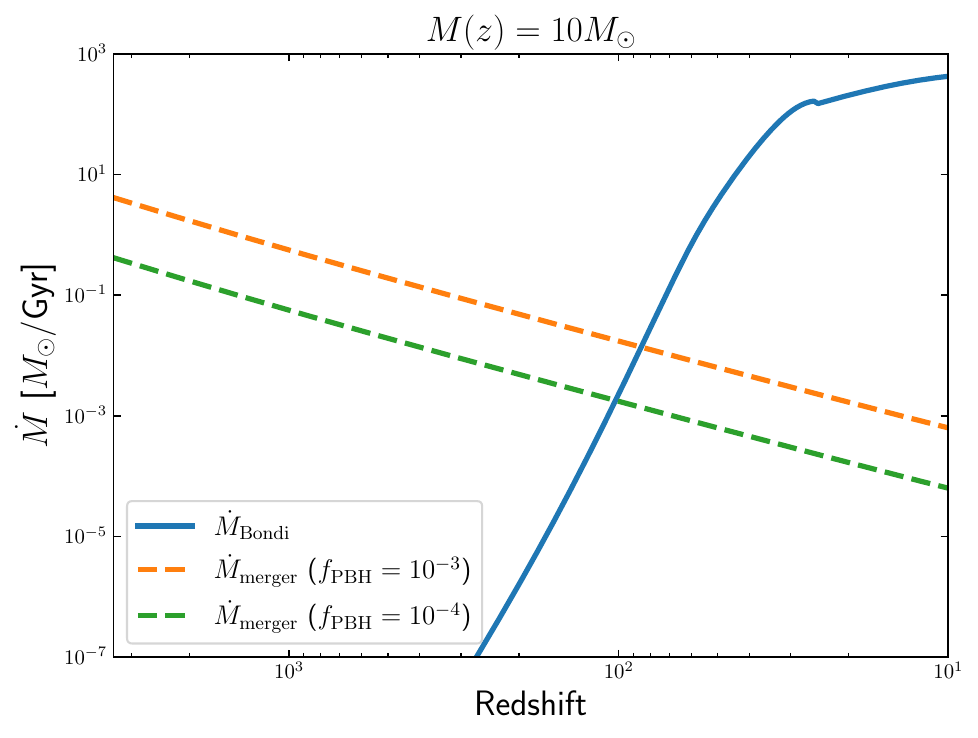}}
    \quad
    {\includegraphics[width=8.5cm]{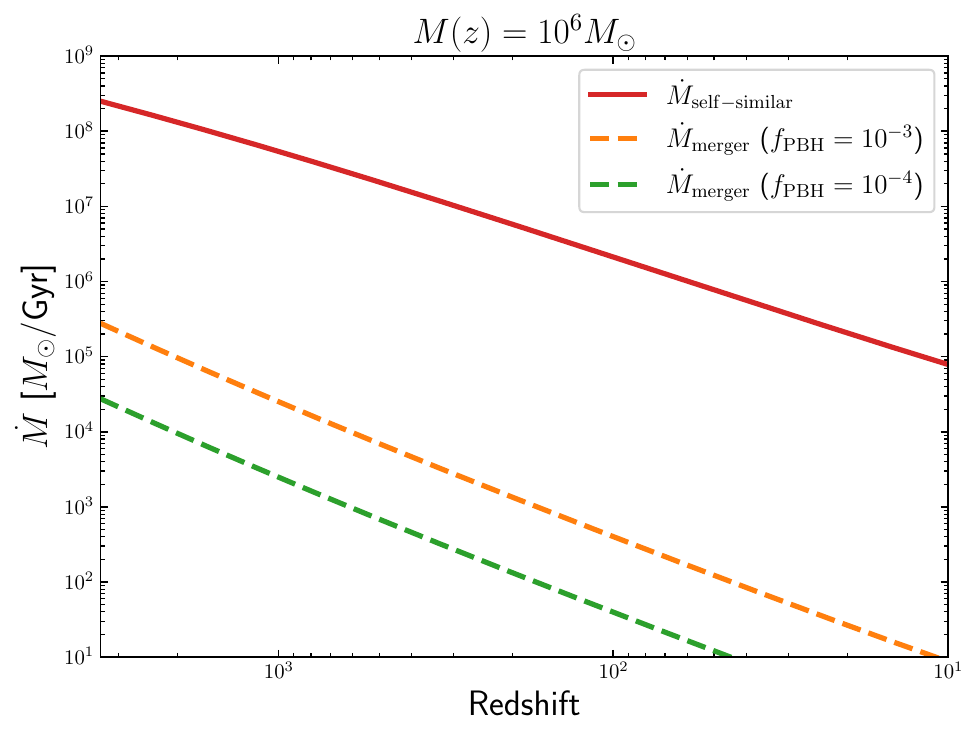}}
\caption{{\it Left panel:} The mass growth rate of a PBH with the
mass of $10~{\rm M}_\odot$ at different redshifts through Bondi
accretion (Equ.~\eqref{eq:bondi}, solid line) and mergers
(Equ.~\eqref{eq:dotmerger_mono}, dotted lines), respectively. The
value of $\dot{M}_{\rm merger}$ depends on the number density of
PBHs, so we assume $f_{\rm PBH}=10^{-3}$ (orange) and $10^{-4}$
(green), respectively.  {\it Right panel:} Same as the left panel
is for the PBH with the mass $10^6{\rm~M}_\odot$, where the
accretion of PBHs is described by self-similar solutions
Equ.~\eqref{eq:self_similar}.}
    \label{fig:accr_vs_merger}
\end{figure*}

In Fig.~\ref{fig:accr_vs_merger}, we see that
for a PBH with $10~{\rm M}_\odot$ mass, its mass growth is
dominated by Bondi accretion only at the redshift
$z\sim\mathcal{O}(10^2)$, depending on the value of $f_{\rm PBH}$,
while for SMPBHs, their mass growth is always dominated by
self-similar accretion due to their tiny number density.
Therefore, we can ignore the effect of
mergers on the mass growth of PBHs safely.

\section{Accretion physics of the steady Bondi solutions}
\label{sec:appendixA}


In this section we provide the basic formalism that
is needed to estimate the Bondi accretion rate onto
PBHs from Equ.~\eqref{eq:bondi}. The mean cosmic gas
density is
\begin{equation} \label{eq:rhogas}
    \rho_{\text{gas}}\simeq250\mu m_p\left(\frac{1+z}
    {1000}\right)^3~\textrm{g}~\textrm{cm}^{-3}\mbox{\ ,}
\end{equation}
and the sound speed of the gas in equilibrium at
the temperature of the intergalactic medium is
\citep{2008ApJ...680..829R,2020JCAP...04..052D,2023arXiv231214097D}
\begin{equation}
    c_s\simeq5.7\left(\frac{1+z}{1000}\right)^{1/2}\left[
    \left(\frac{1+z_{\text{dec}}}{1+z}\right)^\beta+1\right]
    ^{-1/2\beta}~\textrm{km}~\textrm{s}^{-1}\mbox{\ ,}
\end{equation}
with $\beta= 1.72$, and $z_{\text{dec}}\simeq130$ being
the redshift at which the baryonic matter decouples from
the radiation fluid, neglecting the feedback effect induced
by accreting PBHs on the gas.

The accretion parameter $\lambda$ keeps into account of
the Hubble expansion, the coupling of the CMB radiation
to the gas through Compton scattering and the gas viscosity
\citep{2007ApJ...662...53R}. Its analytical expression is
given by \citep{2017PhRvD..95d3534A}
\begin{equation}
    \lambda=\frac{1}{\lambda_{\text{iso}}}\left[\lambda_
    {\text{ad}}+(\lambda_{\text{iso}}-\lambda_{\text{ad}})
    \left(\frac{\gamma_c^2}{88+\gamma_c^2}\right)^{0.22}
    \right]\times\left[\frac{1}{(\sqrt{1+\beta_c}+1)^2}\exp
    \left(\frac{9/2}{3+\beta_c^{3/4}}\right)\right]\mbox{\ ,}
\end{equation}
where $\lambda_{\text{ad}}\approx0.12$ and $\lambda_{\text{iso}}
\approx1.12$. The coefficients $\beta_c$ and $\gamma_c$ are the
dimensionless Compton drag and cooling rates from the surrounding
CMB photons respectively given by \citep{2017PhRvD..95d3534A}
\begin{equation}
    \beta_c=\frac{4}{3}\frac{\chi_e\sigma_T\rho_{\text{cmb}}}
    {m_pc}\frac{GM}{v_{\text{eff}}^3}, \quad \gamma_c=\frac{8}
    {3}\frac{\chi_e\sigma_T\rho_{\text{cmb}}}{m_ec(1+\chi_e)}
    \frac{GM}{v_{\text{eff}}^3}=\frac{2m_p}{m_e(1+\chi_e)}
    \beta_c\mbox{\ .}
\end{equation}
Taking Hubble expansion into account, the coefficient
$\beta_c$ is given by
\begin{equation}
    \beta_c=\left(\frac{M}{10^4~\rm M_\odot}\right)\left(
    \frac{1+z}{1000}\right)^{3/2}\left(\frac{v_{\rm eff}}
    {5.74~{\rm km~s^{-1}}}\right)^{-3}\left[0.257+1.45
    \left(\frac{\chi_e}{0.01}\right)\left(\frac{1+z}{1000}
    \right)^{5/2}\right].
\end{equation}

In the linear regime, the Silk damping acts suppressing
the growth of inhomogeneities on small scales before
the decoupling redshift $z_{\text{dec}}$, such that the
PBH peculiar velocity is of order of the gas sound speed;
for $z<z_{\text{dec}}$, the gas flow lags behind the DM
with a relative velocity $v_{\text{L}}$ (assuming PBHs to
behavee like DM particles \citep{2008ApJ...680..829R}).
\citet{2014PhRvD..89b3519D} explicitly compute the
square root of its variance $\langle v_{\text{L}}^2\rangle$
as a function of time:
\begin{equation}
    \sqrt{\langle v_{\text{L}}^2 \rangle} \approx\min\left[1,
    \frac{1+z}{1000}\right]\times30~\textrm{km}~\textrm{s}^
    {-1}\mbox{\ .}
\end{equation}
Following \citet{2017PhRvD..95d3534A}, we adopt that
\begin{equation}
    v_{\text{eff}}\simeq\sqrt{c_s\sqrt{\langle v_{\text{L}}^2
    \rangle}}\mbox{\ .}
\end{equation}

PBHs with masses larger than $\mathcal{O}({\rm M}_\odot)$ can
comprise only a fraction of the dark matter (DM) in the Universe
according to the current observational constraints, see e.g.
\citet{2021RPPh...84k6902C,2024PhR..1054....1C}. In
corresponding scenario, PBHs would gravitationally attract
surrounding DM particles, leading to an extended DM halo around
PBH. Although direct accretion of DM onto PBHs is negligible, the
halo serves as a catalyst, significantly enhancing the baryonic
accretion rate by several orders of magnitude
\citep{2007ApJ...662...53R,2017JHEAp..13...22R}.

In the presence of an additional DM component, a DM halo
forms during the matter-dominated epoch with mass of
\citep{2007ApJ...665.1277M,2019PhRvD.100b3506A,
2013JCAP...11..059B,2021JCAP...08..053B}
\begin{equation}
    M_h(z)=3M\left(\frac{1+z}{1000}\right)^{-1}\mbox{\ .}
\end{equation}
This is nothing but Equ.~\eqref{eq:35}. Such a halo is
characterised by a typical shperical density profile $\rho
\propto r^{-\alpha}$ with approximately $\alpha\simeq2.25$
\citep{2007ApJ...665.1277M,2019PhRvD.100b3506A}, truncated
at a radius $r_h$.
The presence of a DM halo clothing is usually taken into
account in the accretion parameter $\lambda$
\citep{2007ApJ...662...53R}. To estimate the hierarchy of
scales in the problem, one can define the parameter
\begin{equation}
    \kappa\equiv\frac{r_B}{r_h}=0.22\left(\frac{1+z}{1000}
    \right)\left(\frac{M_h}{\rm~M_\odot}\right)^{2/3}\left(
    \frac{v_{\text{eff}}}{\textrm{km}~\textrm{s}^{-1}}\right)
    ^{-2}\mbox{\ .}
\end{equation}
When $\kappa\ge2$, i.e. when the typical size of the halo
is smaller than the Bondi radius, the accretion rate is
the same as the one for a PBH of point mass $M_h$. On the
other hand, if $\kappa<2$, only a fraction of the dark halo
would be relevant for accretion, and one has to correct the
quantities entering in the parameter $\lambda$
with respect to the naked case as \citep{2007ApJ...662...53R}
\begin{equation}
    \beta_c^{(h)}\equiv\kappa^{\frac{p}{1-p}}\beta_c, \quad
    \lambda^{(h)}\equiv\Upsilon^{\frac{p}{1-p}}\lambda
    (\beta_c^{(h)}), \quad
    \Upsilon=\left(1+10\beta_c^{(h)}\right)^{1/10}
    \exp(2-\kappa)\left(\frac{\kappa}
    {2}\right)^2\mbox{\ ,}
\end{equation}
where $p=3-\alpha\simeq0.75$.

\section{Merger rate of SMPBH binaries}
\label{sec:appendixB}

The merger rate of PBH binaries has been extensively studied,
see e.g. \citet{2016PhRvL.117f1101S,2017PhRvD..96l3523A,
2018ApJ...864...61C,2019PhRvD..99f3523L,2018ApJ...854...41K,
2019JCAP...02..018R,2024arXiv240408416R}. Here, we adopt
the result of \citet{2024PhRvD.109f3515H,2024arXiv240313278H}
for SMPBH binaries, namely
\begin{equation}
    \mathcal{R}(m_i,m_j,t)\approx\frac{1.02\times10^8}
    {\rm Gpc^3yr}f^2\left(\frac{m_i}{\rm~M_\odot}\right)^{-1}
    \left(\frac{m_j}{\rm~M_\odot}\right)^{-1}\left(\frac{m_i+m_j}
    {\rm~M_\odot}\right)\left(\frac{t}{t_0}\right)^{-1}Y
    (y(m_i,m_j,t))\psi(m_i)\psi(m_j),
\end{equation}
where
\begin{equation}
    Y(y)=\frac{21}{2146(1+y^2)}\left[95{}_2F_1\left(-\frac{1}{2},
    \frac{29}{37},\frac{66}{37},-y^2\right)-(58+21y^2){}_2F_1
    \left(\frac{1}{2},\frac{29}{37},\frac{66}{37},-y^2\right)\right],
\end{equation}
with the generalized hypergeometric function ${}_2F_1$ and
the dimensionless quantity
\begin{equation}
    y(m_i,m_j,t)\approx2.95\times10^2f\left(1+\frac{\sigma_
    {\rm eq}^2}{f^2}\right)^{\frac{1}{2}}\left(\frac{m_i}
    {\rm~M_\odot}\right)^{-\frac{1}{7}}\left(\frac{m_j}{\rm~M_\odot}
    \right)^{-\frac{1}{7}}\left(\frac{m_i+m_j}{\rm~M_\odot}\right)
    ^{\frac{1}{21}}\left(\frac{t}{t_0}\right)^{-\frac{1}{7}}.
\end{equation}
Here, $f\approx0.85f_{\rm PBH}$ is the total abundance of
PBHs in nonrelativistic matter, $t_0$ is the present time
and $\sigma_{\rm eq}^2$ is the variance of density perturbations
of the rest of DM at $z_{\rm eq}$.


\bibliography{ms}{}
\bibliographystyle{aasjournal}



\end{document}